\documentclass{IEEEtran}
\usepackage[utf8]{inputenc}
\IEEEoverridecommandlockouts
\usepackage{amsmath,amssymb,amsfonts}
\usepackage{graphicx}
\usepackage[ruled]{algorithm2e}
\usepackage{multirow, multicol}
\usepackage{xcolor}
\usepackage[bookmarks=false, hidelinks]{hyperref}
\usepackage[normalem]{ulem}
\usepackage{algpseudocode}
\usepackage[labelformat=simple]{subcaption}
\usepackage{stmaryrd}

\title{Fast Polarisation-Aware Decoder for Non-Binary Polar Codes}
\author{%
Joseph Jabour, Ali Chamas Al-Ghouwayel, and Emmanuel Boutillon
\thanks{Joseph Jabour is with Lab-STICC, UMR CNRS 6285, IMT Atlantique, Brest, France. (email: \href{mailto:joseph.jabour@imt-atlantique.fr}{joseph.jabour@imt-atlantique.fr})}
\thanks{Ali Al-Ghouwayel is with  Lab-STICC, UMR CNRS 6285, IMT Atlantique, Brest, France. (email: \href{mailto:ali.al-ghouwayel@imt-atlantique.fr}{ali.al-ghouwayel@imt-atlantique.fr})} 
\thanks{Emmanuel Boutillon is with Lab-STICC, UMR CNRS 6285, Université Bretagne Sud, Lorient, France. (email: \href{mailto:emmanuel.boutillon@univ-ubs.fr}{emmanuel.boutillon@univ-ubs.fr})}

\thanks{This work has been funded by the French ANR under grant number ANR-19-CE25-0013-01 and the Lebanese International University. Special thanks to C\'edric Marchand for his indirect but fruitful contribution to this work.}
}
\usepackage[acronym]{glossaries}
\newacronym{LLR}{LLR}{log-likelihood ratio}
\newacronym{FSC-PA}{FSC-PA}{fast successive cancellation polarisation-aware}
\newacronym{BPSK}{BPSK}{binary phase shift keying}
\newacronym{CCSK}{CCSK}{cyclic-code shift keying}
\newacronym{MS}{MS}{min-sum}
\newacronym{EMS}{EMS}{extended min-sum}
\newacronym{AEMS}{AEMS}{asymmetrical extended min-sum}
\newacronym{SC}{SC}{successive cancellation}
\newacronym{NB-SC}{NB-SC}{non-binary successive cancellation}
\newacronym{CA-FSCL32}{CA-FSCL32}{CRC-aided fast successive cancellation list of size $L=32$}
\newacronym{LDPC}{LDPC}{low-density parity check}
\newacronym{NB-LDPC}{NB-LDPC}{non-binary low-density parity check}
\newacronym{NB}{NB}{non-binary}
\newacronym{NB-PC}{NB-PC}{non-binary polar code}
\newacronym{PC}{PC}{polar code}
\newacronym{GF}{GF}{Galois field}
\newacronym{CSS}{CSS}{chip spectrum spreading}
\newacronym{LoRaWan}{LoRaWan}{long range wireless area network}
\newacronym{CN}{CN}{check node}
\newacronym{VN}{VN}{variable node}
\newacronym{AWGN}{AWGN}{additive white Gaussian noise}
\newacronym{PN}{PN}{pseudo-random noise}
\newacronym{SNR}{SNR}{signal-to-noise ratio}
\newacronym{FER}{FER}{frame error rate}

\begin{document}
\maketitle
\begin{abstract}
The paper explores the emerging field of low-complexity \gls{NB} polar decoders. It demonstrates that customising each kernel of a \gls{NB} \gls{PC} through an offline analysis can lead to a significant reduction in the decoder's complexity. The proposed decoder, termed the \gls{FSC-PA}, achieves this by determining the minimum computational load of the parity check nodes that share the same level of input polarisation. The \gls{NB} polar decoder is designed for both \gls{BPSK} and \gls{CCSK} modulations.

The \gls{FSC-PA} algorithm results in an overall reduction of 60\% in field additions and a reduction of 30\% in real additions compared to the state-of-the-art extended min-sum algorithm, with a negligible impact on performance ( with less than 0.2 dB degradation).
\end{abstract}

\glsresetall 
\section{Introduction}
Recent papers highlight the interest in coupling \gls{NB} codes with orthogonal modulations as an effective solution to send short messages with low-rate spectral efficiency (long-range application). In \cite{Liva17}, \gls{NB-LDPC} coded modulation schemes are presented. This scheme exhibits a gain of 0.8 dB compared to state-of-the-art solutions. In \cite{Bourduge23}, the interest of such schemes is also highlighted; however, the complexity of \gls{NB} decoders is circumvented by using multilevel coding with small cardinality \gls{NB} codes at the cost of a small performance penalty. However, having a \gls{NB} code that is easy to decode on a high cardinality \gls{GF} would open many new opportunities to construct an effective short-frame message for long-range transmission. 

\Glspl{PC} are the latest capacity-approaching codes introduced by Arikan \cite{arikanpc} in 2009. The binary \gls{PC} has been adopted in different standards, such as the fifth generation technology standard for broadband cellular networks (5G). Under similar decoding algorithm and coding rate, \glspl{NB-PC} often exhibit superior decoding performance compared to their binary counterparts, particularly in short-packet transmission scenarios. Additionally, \glspl{NB-PC} can be associated with an orthogonal modulation of the same cardinality as the \gls{NB} code. In this paper, \gls{CCSK} modulation is praised for its advantages in terms of demodulation and synchronisation \cite{ccsk_nbcodes, ks_ccsk}. However, other orthogonal modulations, such as the \gls{CSS} modulation used in \gls{LoRaWan} \cite{LoRa} would give identical performance.

The \gls{SC} based decoding suffers from high latency due to the recursive nature of the decoding process. Thus, a fast decoding approach has been proposed in \cite{fast_decoder} to reduce the computational complexity of the binary \gls{PC} decoder and increase its throughput. In \cite{gen_fast_decoder}, a generalised approach to fast decoding of binary \glspl{PC} is introduced to further reduce the decoding latency of the \gls{SC} decoder. The idea is to prune some of the decoder's nodes and reduce the message traversal by performing non-recursive operations based on soft values inputted into these specific nodes. These substitutes are termed in the literature as constituent codes. 
In this work, the constituent codes adopted for fast decoding are Rate-0, Rate-1, and repetition (REP) codes for \gls{NB-PC} decoders. In addition to that, the single-parity check constituent code is adopted for the binary \gls{PC} decoder.

 A concrete design methodology is proposed in \cite{nbpolar_ccsk} for \glspl{NB-PC} along with their kernel coefficients.

The work in \cite{sc_ms} introduces a \gls{SC} decoder based on the \gls{MS} approximation, which operates in the \gls{LLR} domain. The logarithmic representation of the beliefs replaces all multiplication operations with additions. Furthermore, the authors in \cite{nbpolar_ccsk_coeff} show that selecting unitary kernel coefficients maintains the decoder's performance, particularly in the context of CCSK modulation; thus omitting all \gls{GF} multiplication operations. In addition, a restricted set of input messages is proposed in the kernel's \glspl{CN} to reduce decoding complexity with minimal performance impact compared to the conventional \gls{MS} decoding algorithm.

To further simplify the decoding algorithm, \cite{sc_ems} employs the \gls{EMS} algorithm used for \gls{NB-LDPC} decoders \cite{ems_ldpc} to reduce the complexity of the \gls{SC} decoder without sacrificing performance. This approach is later denoted FSC-EMS. Moreover, \gls{SC} decoding with \gls{AEMS} \cite{sc_aems} introduces an asymmetric approach to processing the CN inputs for output generation and is denoted as FSC-AEMS in the sequel. This decoding approach leverages the efficient L-bubble algorithm in \cite{l_bubble} to process the reduced input set within the \gls{CN}. This optimised implementation ensures streamlined and effective CN processing with a decoding performance similar to the conventional FSC-EMS decoder when associated with \gls{CCSK} modulation. However, using the L-bubble sorter with asymmetric processing degrades the decoding performance under \gls{BPSK} modulation.

This paper proposes a practical optimisation based on statistical estimates to reduce the complexity of the \gls{NB-PC} decoder. The optimisation observes the polarisation effect on the kernels and utilises the statistical information to generate a polarisation-aware decoder that only computes the information that is likely to be useful. This proposed optimisation approach can be applied naturally to the fast decoding algorithms proposed in \cite{fast_decoder, gen_fast_decoder} to allow further reduction of the computational complexity of the decoder, by shortening the sizes of the messages entering the nodes in the upper layers of the \gls{NB-PC} decoder. Such constituent nodes, which kept their initial recursive processing of message traversal unchanged, strongly dominate the global complexity of the NB-PC decoder.

The paper is outlined as follows: Section \ref{sec:polar_code} introduces the \glspl{PC} and the SC decoder. Section \ref{sec:methodology} discusses the \gls{FSC-PA}. In Section \ref{sec:complexity}, the complexity and simulation results of \gls{FSC-PA} decoders are compared with the state-of-the-art \gls{SC} decoders. Lastly, the main points are concluded in Section \ref{sec:conc}.

\section{Non-Binary Polar Codes and Decoders}\label{sec:polar_code}
An \gls{NB-PC} is a \gls{PC} defined over GF$(q=2^p)$ with $p>1$. The structure of a \gls{PC} depends on the adopted kernel. A kernel $G_2$ is a generator matrix for $N=2$ and is considered the basic transformation unit. The adopted kernel $G_2$ transforms an input $u=(u_0, u_1)$ into an output $x=(x_0, x_1)$ using the following relation.
\begin{equation}
    \begin{cases}
    x_0=u_0 \oplus u_1\\
    x_1=\gamma \circledast  u_1,
    \end{cases}
\end{equation}
where $\circledast $ represents the field multiplication and $\oplus$ represents the field addition over GF$(q)$, and $\gamma \in$ GF$(q)$.

The polar encoder generates a codeword based on the polar transformation that leads to channel polarisation. The encoder encodes the message $m$ of size $K$ into a codeword $c$ of size $N=2^n$ by allocating the message $m$ into the $K$ most reliable channel positions (having the least error probability $\mathcal{P}_{\epsilon}$) such that $u_i=m_j: i\in \mathcal{A}_D, 0\leq j \leq K$ where $\mathcal{A}_D$ represents the set of the $K$ most reliable input positions (noiseless virtual channels). The remaining $N-K$ noisiest channels are frozen so that $u_i$ is equal to the null element of GF$(q) \ \forall \ i \notin \mathcal{A}_D$. The frozen symbol channels are determined by estimating the error rate in each channel using a genie-aided decoder \cite{arikanpc}.

Subsequently, the message $u$ is transformed into the codeword $x$ using a generator matrix $G_N$ generated as the $n^{th}$ Kronecker power of $G_2$, i.e., $G_N=G_2^{\otimes n}$, with $\otimes$ representing the Kronecker product. 

On the receiver side, the \gls{SC} decoder generates an estimate $\hat{u}$ of the information message $u$ from the received noisy codeword $y$. The input vectors of the polar decoder $L^{(0)}_{i}$ are the intrinsic \gls{LLR} information observed from the channel and can generally be expressed as
\begin{multline}
    \label{eq:llr_general}
    L^{(0)}_i(\alpha(k))= \log \mathbb{P}(y_i|\hat{\alpha})- \log \mathbb{P}(y_i|\alpha), \\ \forall \ i \in \llbracket0, N-1\rrbracket, \alpha \in \text{GF}(q),
\end{multline}
where $\hat{\alpha}=\underset{\alpha \in GF(q)}{\arg\max{\mathbb{P}(y_i|\alpha)}}$, that is, $\hat{\alpha}$ is the hard decision.

\subsection{Log-Likelihood Ratio Estimation in \gls{BPSK} Modulation}
The intrinsic \gls{LLR} input  $L^{(0)}_i \ \forall i \in \llbracket 0, N-1\rrbracket$ of a binary polar decoder can be deduced in the context of BPSK modulation over binary \gls{AWGN} channel as 
\begin{equation}\label{eq:llr_compute_bpsk_binary}
    \Lambda^{(0)}_i =  \frac{2y_i}{\sigma^2},
\end{equation}
where $\sigma^2$ denotes the noise variance.

For \gls{NB} polar decoding, each $p$-bits correspond to a given symbol $\alpha$. An input \gls{LLR} $ L^{(0)}_i$ is a vector of $q$ corresponding to the reliability of each possible symbol $\alpha \in GF(q)$. Thus, can be expressed as
\begin{equation}\label{eq:llr_compute_bpsk}
    L^{(0)}_i(\alpha) =  \sum_{k=0}^{p-1}\frac{y_i(k)}{\sigma^2}(\hat{\alpha}(k) - \alpha(k)) \ \forall \ \alpha \in \text{GF}(q),
\end{equation}
where $\alpha(k)$ denotes the $k^\text{th}$ bit in the binary representation of the symbol $\alpha \in \text{GF}(q)$.

\subsection{Log-Likelihood Ratio Estimation in \gls{CCSK} Modulation}
On the other hand, \gls{CCSK} modulation \cite{ccsk} is a spread spectrum modulation that spreads an encoded symbol of $p-$bits into a modulated symbol of $q=2^p$ bits (chips) using a \gls{PN} sequence. Let $\psi=(\psi_{x_0}, \psi_{x_1}, \cdots, \psi_{x_{N-1}})$ denote the \gls{CCSK} modulated frame of a codeword $x=(x_0, \cdots, x_{N-1})$ where $x_i \in \text{GF}(q)$. The $i^{th}$ transmitted \gls{CCSK} symbol $\psi_{x_i}$ is received as a vector $y_i = (y_i(k))_{k=0,1,\ldots, q-1}$ with $y_i(k) = (2\psi_{x_i}(k)-1) + w_i(k)$ where $w_i(k)$ represents a sample of real additive noise. Therefore, the intrinsic \gls{LLR} vectors $L^{(0)}_i$ can be expressed in CCSK modulation over an \gls{AWGN} channel as 
\begin{equation}\label{eq:llr_compute_ccsk}
    L^{(0)}_i(\alpha) =  \sum_{k=0}^{q-1}\frac{y_i(k)}{\sigma^2}(\psi_{\hat{\alpha}_i}(k) - \psi_{\alpha}(k)) \ \forall \ \alpha \in \text{GF}(q),
\end{equation}
where $\hat{\alpha}_i$ represents the maximum likelihood decision on $y_i$ and defined as 
\begin{equation}\label{eq:llr_ml}
    \hat{\alpha}_i = \underset{\alpha \in\ GF(q)}{\mathrm{argmax}} \{\sum_{k=0}^{q-1}\frac{y_i(k)}{\sigma^2}\psi_{\alpha}(k)\}.
\end{equation}
Note that, by construction, the \gls{LLR} values are all positive and $L^{(0)}_i(\hat{\alpha}_i)=0$.

For binary decoding over \gls{CCSK} modulation, binary marginalisation is performed to determine the bit-level reliabilities from the symbol reliabilities. For each symbol $x_i$, the intrinsic LLR vector $L^{(0)}_i(\alpha)$ is obtained first in $\text{GF}(q)$. Then, for each bit position $k \in {0, 1, ..., p-1}$ of the symbol’s binary representation, a corresponding binary LLR using is approximated as
\begin{equation}
\label{eq:binary_marginalization}
% \Lambda_i^{(k)} = \log \frac{\sum\limits_{\alpha \in \text{GF}(q): \alpha(k)=0} \exp(-L^{(0)}_i(\alpha))}{\sum\limits_{\alpha \in \text{GF}(q): \alpha(k)=1} \exp(-L^{(0)}_i(\alpha))},
\Lambda_{i}^{(0)}(k) = \max\limits_{\alpha_0 \in \text{GF}(q): \alpha_0(k)=0}{L^{(0)}_i(\alpha_0)} - \max\limits_{\alpha_1 \in \text{GF}(q): \alpha_1(k)=1} {L^{(0)}_i(\alpha_1)}.
\end{equation}

\subsection{Polar Successive Cancellation Decoder}
\begin{figure}[t]
    \centering
    \includegraphics[width=\columnwidth]{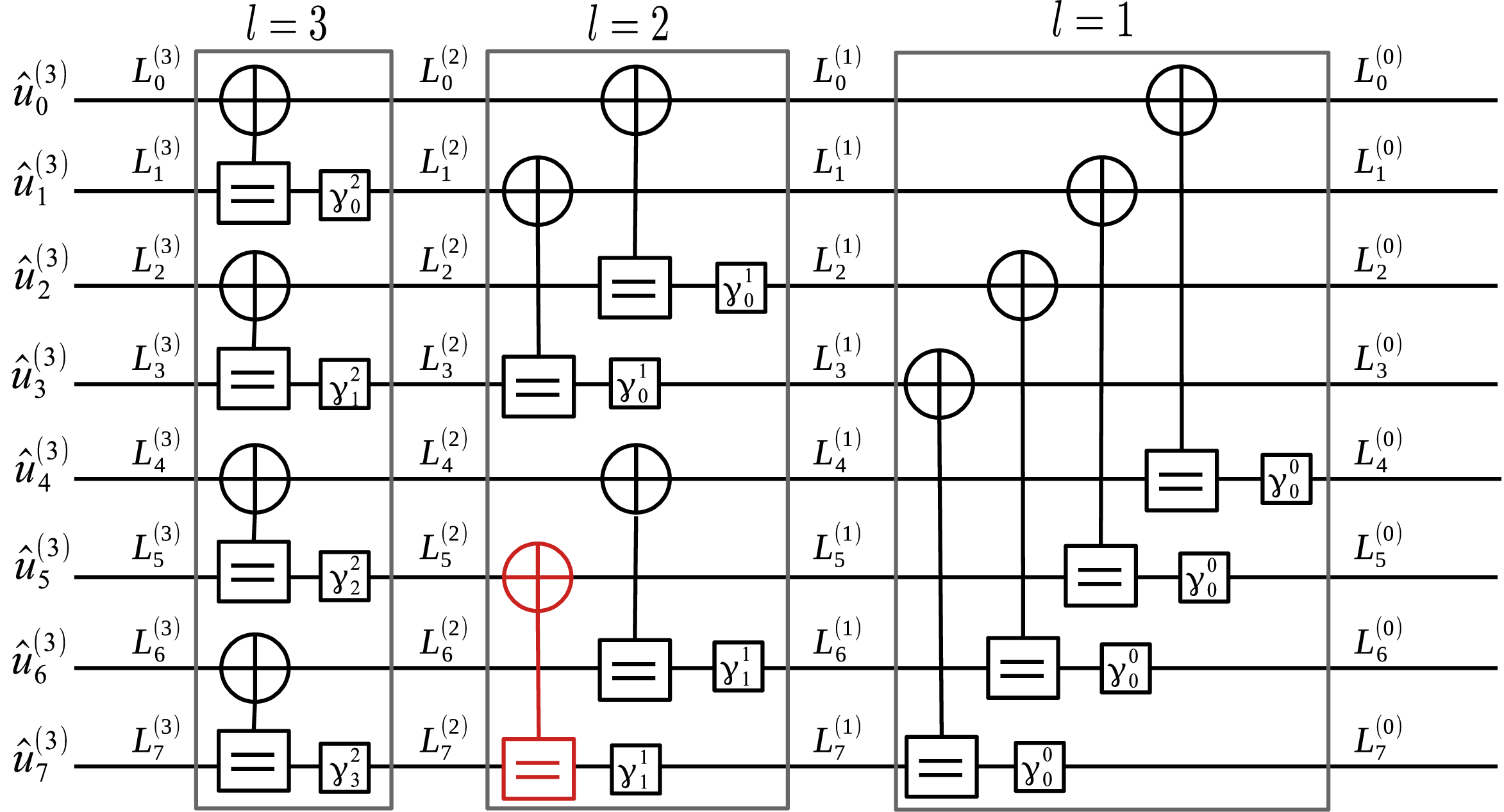}
    \caption{SC Polar Decoder for $N=8$.}
    \label{fig:decoderN8}
\end{figure}

A polar decoder consists of $n=\log_2(N)$ layers, where each layer has $N/2$ kernels, as shown in Fig.~\ref{fig:decoderN8}. Each kernel comprises a single-parity check node indicated simply as \gls{CN} and a repetition node called a \gls{VN}.

A kernel at instance $t$ of layer $l$ has two input vectors denoted as $L_{\theta_t^{l-1}}^{(l-1)}$ and $L_{\phi_t^{l-1}}^{(l-1)}$ at indices $\theta_t^{l-1}$ and $\phi_t^{l-1}$ deduced as follows:
\begin{equation}
    \begin{gathered}
    \theta_t^{l-1}=2t-(t \bmod 2^{n-l})\\
    \phi_t^{l-1}=2^{n-l}+2t-(t \bmod 2^{n-l}),    
    \end{gathered}
    \label{eq:input_index}
\end{equation}
for $t=0,\dots, N/2-1$. The kernel instance $t$ and the layer $l$ are omitted from the indices $\theta_t^{l-1}$ and $\phi_t^{l-1}$ in the sequel for better readability of the notation. As an example, the red kernel presented in Fig.~\ref{fig:decoderN8} is in layer $l=2$ and instance $t=3$. Therefore, according to \eqref{eq:input_index}, the corresponding \gls{LLR} inputs $L_{\theta_t^{l-1}}^{(l-1)}$ and $L_{\phi_t^{l-1}}^{(l-1)}$ are the vectors with indices $\theta_3^{1}=5$ and $\phi_3^{1}=7$, respectively, that is, the corresponding \gls{LLR} input vectors of the kernel at instance $t=3$ and the layer $l=2$ are $L_{5}^{(1)}$ and $L_{7}^{(1)}$.

\begin{figure}[t]
    \centering
    \includegraphics[width=0.8\columnwidth]{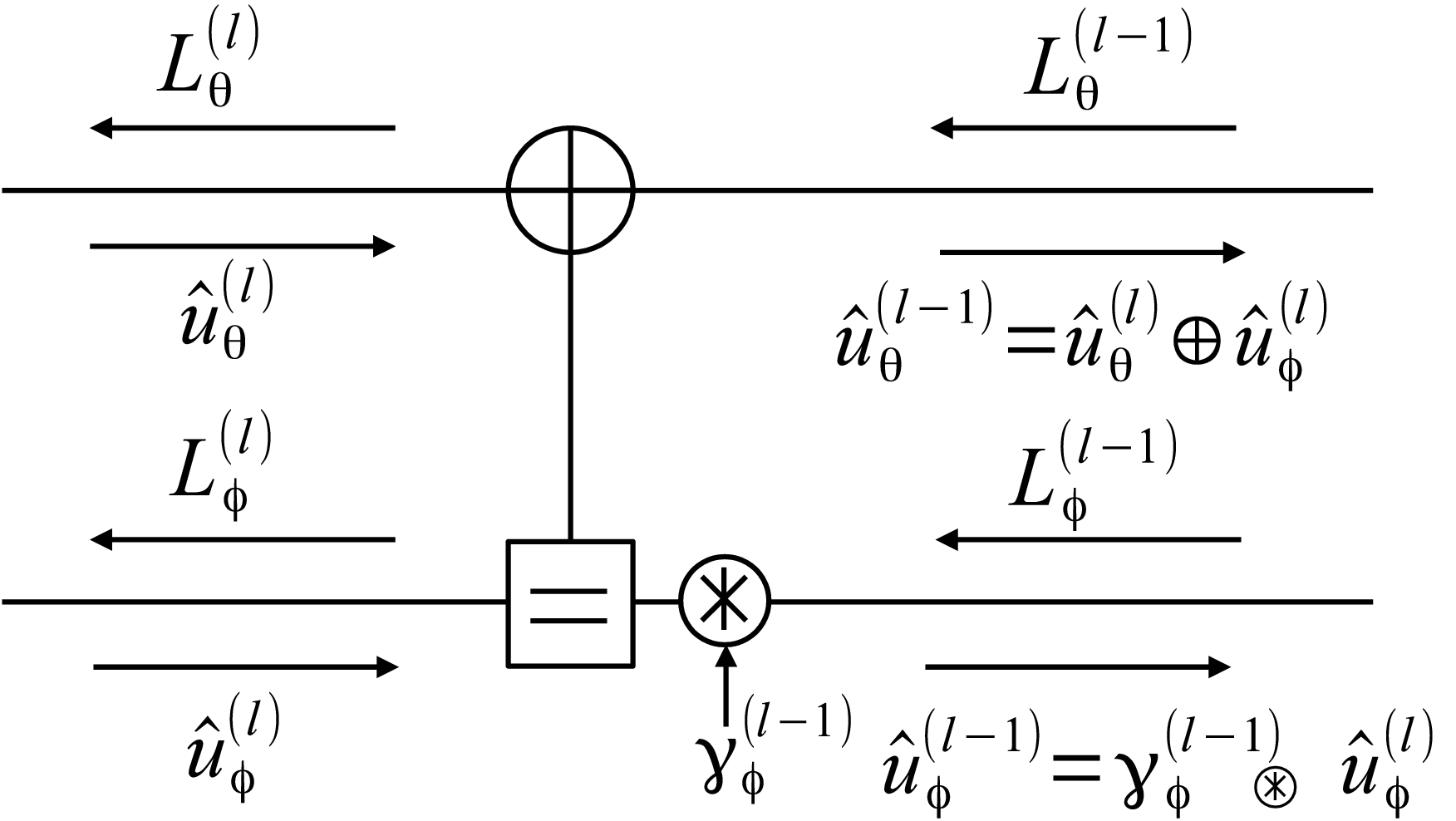}
    \caption{Kernel Nodes Processing }
    \label{fig:kernel_nodes}
\end{figure}

The inputs $L_\theta^{(l-1)}$ and $L_\phi^{(l-1)}$ are processed by the kernel to obtain the output vectors $L_\theta^{(l)}$ and $L_\phi^{(l)}$ using the \gls{CN} and \gls{VN} update processing, respectively, as illustrated in Fig.~\ref{fig:kernel_nodes}.

The \gls{CN} update of a kernel can be expressed using the \gls{MS} approximation as
\begin{multline}
   L_\theta^{(l)}(\beta)=\min_{{\eta \in \text{GF}(q)}}{(L_\theta^{(l-1)}(\beta \oplus \eta) +L_\phi^{(l-1)}(\gamma \circledast  \eta))}, \\
   \forall \ \beta \in \text{GF}(q).
    \label{eq:cn_update_ms}
\end{multline}

Similarly, the \gls{VN} update of a kernel can be expressed as
\begin{equation}
       L^{(l)}_\phi(\eta)=L_\theta^{(l-1)}(\hat{u}^{(l)}_\theta \oplus \eta)+L_\phi^{(l-1)}(\gamma \circledast  \eta), \ \forall \ \eta \in \text{GF}(q),
       \label{eq:vn_update_ms}
\end{equation}
where $\hat{u}_{\theta}^{(l)}$ is the estimated symbol at index $\theta$ of layer $l$. To avoid arithmetic overflow, the vector $L^{(l)}_\phi$ is normalised as follows.
\begin{equation}
     L^{(l)}_\phi(\eta)= L^{(l)}_\phi(\eta)-\min( L^{(l)}_\phi), \ \forall\  \eta \in \text{GF}(q).
     \label{eq:vn_normalize}
\end{equation}

The hard decision $\hat{u}^{(n)}_i$ at layer $l=n$ and for a node $i=0,1,\dots,N-1$ is estimated as 
\begin{equation}
    \hat{u}^{(n)}_i=
    \begin{cases}
    0,& \text{if } i \notin \mathcal{A}_D\\
    \underset{\alpha \in\ \text{GF}(q)}{\arg \min} \ L^{(n)}_{i}(\alpha), & \text{if } i \in \mathcal{A}_D. 
    \end{cases}
\label{eq:hard_decision}    
\end{equation}

Once computed, the estimated symbols $\hat{u}_{\theta}^{(l)}$ and $\hat{u}_{\phi}^{(l)}$ at layer $l$ are back-propagated to layer $l-1$ as
\begin{equation}
\begin{gathered}
    \hat{u}_{\theta}^{(l-1)}=\hat{u}_{\theta}^{(l)} \oplus \hat{u}_{\phi}^{(l)}, \ \ 
    \hat{u}_{\phi}^{(l-1)}= \gamma \circledast  \hat{u}_{\phi}^{(l)}.
    \end{gathered}
    \label{eq:backward_update}
\end{equation}

\section{Successive Cancellation Polarisation-Aware Decoder}\label{sec:methodology}
The \gls{FSC-PA} decoding algorithm is a simplified decoding approach for \gls{NB-PC}. It is based on investigating the statistical behaviour of the computations performed within the \glspl{CN} at different layers. %\joseph{NEWLY ADDED, IF APPROVED PLEASE UNCOMMENT: The statistical behaviour of the check node processing can be determined by either density evolution and Monte Carlo approaches. Although density evolution offers a theoretically rigorous approach, its application becomes intractably complex for \gls{NB} codes due to the intricate structure of the associated probability densities. Therefore, a Monte Carlo simulation-based method is used to estimate the relevant probabilities, mirroring the approach commonly used in \gls{NB-LDPC} decoding. This pragmatic alternative enables an accurate estimation of the message distributions at each decoding stage without the prohibitive analytical burden associated with density evolution, making it more suitable for practical implementation and performance evaluation.}

In the domain of NB-LDPC codes, equations of density evolution have been derived \cite{rathi2005}. In the case of \gls{NB}-\glspl{PC}, these equations are also applicable. However, two observations prevent us from mathematically formalising the density function of incoming/outgoing messages from the kernels. First, they are difficult to handle in the general case, and Monte Carlo methods are used in practice to evaluate the decoding threshold of NB-LDPC codes. Last but not least, the theoretical analysis of the proposed polarisation-aware algorithm requires knowledge of the probability distribution of the first $n_m$ most reliable decisions of the messages. This problem cannot be tackled simply because it requires manipulation of the distributions over $[0, 1]^{n_m}$. Consequently, Monte Carlo simulation is relied upon in the sequel.

\subsection{Definition of Nodes Clusters}
In a polar decoder of size $N$, there are $N/2$ kernels in each of the $n$ layers, each consisting of a \gls{CN} and a \gls{VN}. Nodes that share the same polarisation level can be clustered \cite{nbpolar_ccsk_coeff} based on their polarisation parameter (mutual information, Bhattacharyya parameter or synthesised channel error rate).

 In a similar pattern, let a cluster of kernels $S_s^{(l)}$ at layer $l$ with index $s=0, \dots, 2^{l-1}-1$ be a set of kernels at instances $t=s\times(2^{n-l})$ up to $t=(s+1)\times 2^{n-l}-1$. Thus, a cluster $S_s^{(l)}$ includes $2^{n-l}$ kernels with all \glspl{CN} having similar polarisation levels and similarly all \glspl{VN} having similar polarisation levels (which differ from those of \glspl{CN}). In this approach, \glspl{VN} of the kernels are considered to belong to the same cluster as \glspl{CN} as they both share the same input messages. Thus, any optimisation at \glspl{CN} will have a similar impact on their corresponding \glspl{VN}.
 
Hence, the cluster definitions serve as a foundation for dynamically computing the kernel message sizes within each cluster, based on statistical estimations at the \glspl{CN}.

\subsection{Check Node Processing Prerequisite}
Assume an \gls{EMS}-based CN processing at layer $l$, and kernel $t$ with the two sorted input messages $M^{(l-1)}_{\theta}$ and $M^{(l-1)}_{\phi}$ (most reliable candidates of $L^{(l-1)}_{\theta}$ and $L^{(l-1)}_{\phi}$ respectively) each of size $n_m$. Each element of the input messages consists of a 2-tuple whose first and second instances represent, respectively, the \gls{GF} symbol and its associated \gls{LLR} value. For clarity, the \gls{GF} vector of the message $M^{(l-1)}_{\theta}$ is denoted as $M^{(l-1)\oplus}_{\theta}$ and the \gls{LLR} vector is denoted as $M^{(l-1)^+}_{\theta}$. %Thus, for any index $i$,  $M^{(l-1)}_i(j) = (M^{(l-1)\oplus}_i(j),M^{(l-1)+}_i(j))$, with $j=0,1,\ldots, n_m - 1$, $M^{(l-1)+}_i(0)= 0$ and $M^{(l-1)+}_i(j) \leq M^{(l-1)+}_i(j')$ if $j<j'$. 

Let the vectors $M_{L}$ and $M_{H}$ correspond to the least reliable input vector and the highest reliable input vector, respectively, of the inputs $M^{(l-1)}_{\theta}$ and $M^{(l-1)}_{\phi}$. Relative reliability \cite{sc_aems} is estimated by comparing an \gls{LLR} value of each input $M^{(l-1)}_{\theta}$ and $M^{(l-1)}_{\phi}$ with an index $z=2$ such that
\begin{equation}
    \begin{gathered}
        (M_{L}, M_{H}) \leftarrow \\
        \begin{cases}
            (M^{(l-1)}_{\theta}, M^{(l-1)}_{\phi}) & \text{ If }(M^{(l-1)}_{\theta}(z)<M^{(l-1)}_{\phi}(z))\\
            ( M^{(l-1)}_{\phi}, M^{(l-1)}_{\theta}) & \text{ Otherwise}.
        \end{cases}
    \end{gathered}
    \label{eq:relocate_inputs}
\end{equation}

Furthermore, let the matrix $T'_\Sigma$ be the intermediate matrix of all potential candidates to generate the output $M_{\theta}^{l}$ such that
\begin{equation}
    \begin{gathered}
            T'_{\Sigma}(i,j)=M_{H}(i) \boxplus M_{L}(j)
            :\ 0 \leq i,j < n_m,
    \end{gathered}
    \label{eq:reallocate_tsigma}
\end{equation}
where $\boxplus$ corresponds to a \gls{GF} $(\oplus)$ and an \gls{LLR} addition $(+)$ operations.

Thus, the most reliable (distinct) $n_m$ symbols ${T'}_{\Sigma}^{\oplus}$ of ${T'}_{\Sigma}$ with their corresponding \glspl{LLR} ${T'}_{\Sigma}^{+}$ are sorted to generate the message $M_\theta^{(l)}$. 

\subsection{Statistical Computation Using EMS-based Check Nodes}
The relative reliability approach defined in \cite{sc_aems}, allows to estimate the potential elements generated in \eqref{eq:reallocate_tsigma}. This helps reduce the overall complexity of the SC decoder and can be achieved by computing the bubble pattern matrices $\mathcal{B}^{(l)}_t$. 

A bubble pattern matrix is defined as a matrix of size $n_m \times n_m$ which includes the probability that the element (bubble) of $T'_\Sigma(i,j)$ is selected in any output element of $M^{(l)}_\theta$ in layer $l$ and CN at kernel instance $t$. To do so, for a decoded codeword, a bubble pattern matrix $\mathcal{B}^{(l)}_t$ can be computed as follows.
\begin{equation}
\begin{gathered}
     \mathcal{B}^{(l)}_t(i,j)=
\begin{cases}
    1 & \text{If } T'_{\Sigma}(i,j) \in \ M^{(l)}_\theta\\
    0 & \text{Otherwise}
\end{cases} ,
 \\ \forall \ 0 \leq t< N/2; \ 1 \leq l < n.
\end{gathered}
\label{eq:bubble_pattern}
\end{equation}

Estimating the average behaviour of the bubble pattern of a given CN gives useful insight into partitioning its bubbles (the element of $T'_\Sigma$) into two sets: the set of useful bubbles (i.e. the ones that are frequently used) and the set of useless bubbles (i.e., the ones that are never, or only rarely used). This partition allows for reducing the complexity of the CN processing by only computing the set of useful bubbles. Experimentation shows that the \glspl{CN} with similar polarisation effects have similar sets of useful bubbles. Thus, the accumulation of statistical information can help increase the accuracy of the average behaviour of bubbles.

  A contribution rate matrix $\mathcal{C}^{(l)}_s$ is defined as the aggregated statistical information $\mathcal{B}^{(l)}_t$ of \glspl{CN} at instances $t\in \mathcal{S}^{(l)}_s$. Hence, the contribution rate matrix $\mathcal{C}^{(l)}_s$ at layer $l$ and cluster $s$ can be expressed as
    \begin{equation}
    \begin{gathered}
         \mathcal{C}^{(l)}_s=\frac{1}{2^{n-l}}\sum_{t=s.2^{n-l}}^{(s+1).2^{n-l}-1}\mathcal{B}^{(l)}_t.
    \end{gathered}
    \label{eq:contribution_rate}
\end{equation}

The bubble pattern matrices $\mathcal{B}^{(l)}_t$ are taken into account if and only if the frame is successfully decoded to avoid biasing the statistical results by a faulty decision. The estimation of $\mathcal{C}^{(l)}_s$ is accumulated for $N_r$ well-decoded frames. The contribution rate matrices are normalised such that  
\begin{equation}
    C^{(l)}_s = \frac{\mathcal{C}^{(l)}_s}{N_r}.
    \label{eq:contribution_rate_norm}
\end{equation}

 The last layer $l=n$ includes simple \glspl{CN} that perform a field addition of the most reliable element of the two inputs to obtain the estimate $\hat{u}_i \ \forall \ i \mod 2 = 0$. Therefore, they are excluded from the simplification process. 

\subsection{Bubbles Pruning Process}
The pruning process is performed offline to obtain a pruned form of $T'_\Sigma$, denoted as $T'^{(l)}_{\Sigma_s}$, for all clusters $s=0, \dots, 2^{l-1}-1$ in layers $l=1, \dots, n-1$. In the pruning process, a threshold $\mathcal{P}_t$ is defined as the minimum contribution rate value for a bubble to be computed. Therefore, any bubble with a contribution rate $C^{(l)}_s(i,j)<\mathcal{P}_t$ is omitted from the \gls{CN} processing within the cluster $\mathcal{S}^{(l)}_s$. 

The indicator matrices $\mathcal{R}^{(l)}_{s}$ can be found for all clusters $s=0, 1, \dots , 2^{l-1}-1$ over the $l=1, 2,\dots, n-1$ layer as follows:
\begin{equation}
    \mathcal{R}^{(l)}_{s}=
    \begin{cases}
        1 & \text{ If } C^{(l)}_s(i,j)>\mathcal{P}_t\\
        0 & \text{Otherwise.}
    \end{cases}
    \label{eq:potential_region}
\end{equation}

These $\mathcal{R}^{(l)}_{s}$ matrices indicate which bubbles of $T'^{(l)}_{\Sigma_s}$ should be exclusively computed to generate $M^{(l)}_\theta$, that is, 

\begin{equation}\label{eq:tsigma_pa}
\begin{gathered}
    T'^{(l)}_{\Sigma_s}(i,j)=
    \begin{cases}
        M_{H}(i) \boxplus M_{L}(j) & \text{If }  \mathcal{R}^{(l)}_{s}(i,j)=1\\
        (0, +\infty) & \text{Otherwise.}
    \end{cases}\\
    : 0\leq i,j < n_{s'}^{(l-1)}.
    \end{gathered}
\end{equation}
where $s' = \lfloor s/2 \rfloor$.

The tuple $(0, \infty)$ represents the \gls{GF} and \gls{LLR} values resulting from the saturation process, in which the element $T'^{(l)}_{\Sigma_s}(i,j)$ is excluded from the output generation.

Moreover, since the bubbles in the first row are generated directly from the corresponding input $M^{(l-1)}_{L}$, the omitted bubbles consequently reduce the maximum size of the propagated messages so that the size of the CN input messages within the node cluster $\mathcal{S}^{(l)}_s$ is
\begin{equation}
    \begin{gathered}
        n^{(l)}_{s}=\sum_{j=0}^{n^{(l-1)}_{s'}}{\mathcal{R}^{(l)}_{s}(0,j)},
    \end{gathered}
    \label{eq:pa_size}
\end{equation}
The size of the inputs at layer $l=1$ denoted as $n_0^{(0)}$ is equal to the chosen \gls{EMS} size $n_m$, i.e., $n_0^{(0)}=n_m$. 

As a result, both the \glspl{CN} and \glspl{VN} in cluster $S^{(l)}_s$ have an output size of $n^{(l)}_s$ elements but different input sizes. A \gls{CN} in a cluster $S_s^{(l)}$ generates an output of $n^{(l)}_s$ elements and has inputs of $n^{(l-1)}_{s'}$ elements. Therefore, only $n^{(l)}_s$ among the $n^{(l-1)}_{s'}$ elements of the inputs are sufficient to generate $n^{(l)}_s$ reliable elements (see summation boundary of \eqref{eq:tsigma_pa}). %The reason behind generating $n^{(l-1)}_{s'}$ elements from the nodes in the previous layer is that the \glspl{VN} belonging to cluster $S_s^{(l)}$ process the $n^{(l-1)}_{s'}$ elements of the corresponding inputs to generate $n^{(l)}_s$ reliable elements at the output. This is due to the processing nature of the \gls{VN} which relies on the common \gls{GF} elements of the inputs to generate reliable (meaningful) output. 
Nonetheless, the reason for generating $n^{(l-1)}_{s'}$ elements in the previous layer is that the \glspl{VN} in cluster $S^{(l)}_s$ require $n^{(l-1)}_{s'}$ input elements to produce $n^{(l)}_s$ reliable outputs. This is due to the processing nature of the \gls{VN} which combines the reliability of the common \gls{GF} elements to generate meaningful results.

\begin{figure}[t]
    \centering
          \begin{subfigure}[b]{0.49\columnwidth}
         \centering
         \includegraphics[width=\columnwidth]{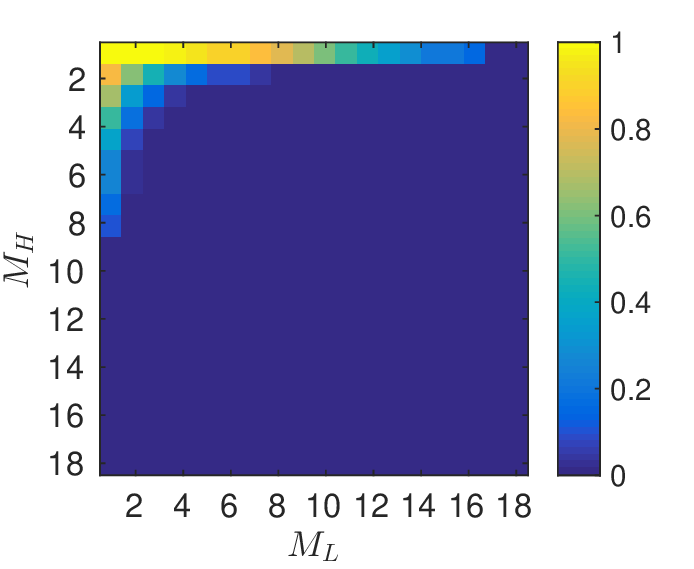}
         \caption{Contribution Rate Matrix $C^{(2)}_{1}$}
         \label{fig:N64K11}
     \end{subfigure}
          \begin{subfigure}[b]{0.49\columnwidth}
         \centering
         \includegraphics[width=\columnwidth]{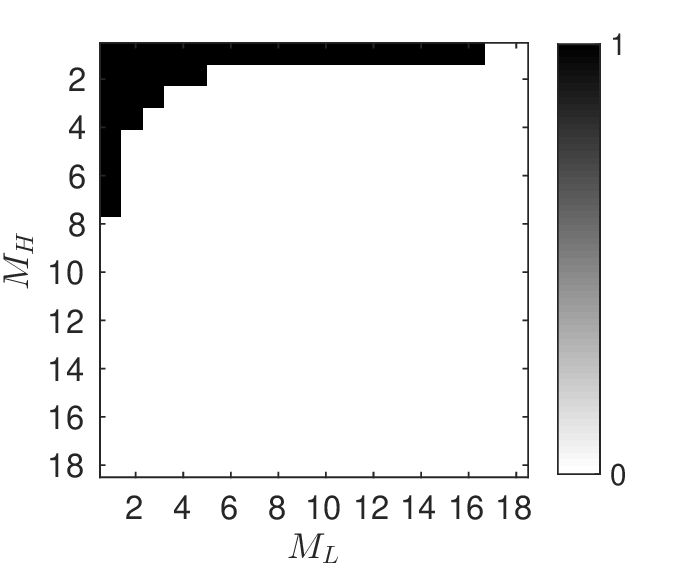}
         \caption{Indicator Matrix $ \mathcal{R}^{(2)}_1$}
             \label{fig:N64K11_2}
     \end{subfigure}
      \caption{Statistics and Pruning at $l=2$ and $s=1$ over $N=64$.}
      \label{fig:contribtuion_rate_example}
\end{figure}

For further clarification, the contribution rate matrix $C^{(2)}_1$ is depicted in Fig.~\ref{fig:N64K11} for a code length of $N=64$, and information length of $K=11$ on GF$(64)$ over \gls{CCSK} modulation. The rate is estimated at a \gls{SNR} of $-13.5$ dB for the cluster $s=1$ of layer $l=2$. In addition, the corresponding indicator matrix $ \mathcal{R}^{(2)}_1$ is demonstrated in Fig.~\ref{fig:N64K11_2} with a threshold of $\mathcal{P}_t=0.12$.

Algorithm~\ref{algo:sc-pa} summarises the design procedure for the proposed \gls{FSC-PA} decoder, based on the approach described above.

\begin{algorithm}\LinesNumbered
    \KwIn{$q, N, K, n_m, N_{r}, \mathcal{P}_t$}
    \textbf{Initialization Step:} \\Set $C^{(l)}_s$ to the $n_m \times n_m$ null matrix, $l=1,2,\ldots, n-1$, $s = 0, 1, \ldots, 2^{l-1}-1$.\\ 
    \textbf{Pre-Processing Step:} Determine SNR $\mu$ required to have a \gls{FER} of $10^{-2}$ using FSC-EMS with $n_m$.\\
    \While{$i<N_{r}$}{
        \textbf{Step 1:} Receive codeword $Y$ via AWGN at SNR $\mu$. \\
        \textbf{Step 2:} Decode $Y$ using FSC-EMS decoder.\\
         \textbf{Step 3:} Trace useful bubbles for statistical analysis\\
                    % 1. Compute $T'_{\Sigma}$ as in \eqref{eq:reallocate_tsigma} with $M^{(l-1)}_{\theta}$ and $M^{(l-1)}_{\phi}$ inputted at indices \eqref{eq:input_index} to generate $M^{(l)}_\theta$.\\
                    % 2. Compute $\mathcal{B}^{(l)}_t$ as in \eqref{eq:bubble_pattern}.
                    Compute \eqref{eq:reallocate_tsigma} and \eqref{eq:bubble_pattern}.\\
             \textbf{Step 4:} Computation of Contribution Rate Matrix\\
             \uIf{decoding success}{
                Compute \eqref{eq:contribution_rate}.
                }
             }
    \textbf{Step 6: } Pruning Process: 
    Normalize $C^{(l)}_s$ as in \eqref{eq:contribution_rate_norm}, deduce $\mathcal{R}^{(l)}_{s}$ and $n_s^{(l)}$ as in \eqref{eq:potential_region} and \eqref{eq:pa_size}.\\
    \KwOut{$\mathcal{R}^{(l)}_s$.} %\forall \ l=\{1, \dots, n-1\}, s=\{0, \dots, 2^{l-1}-1\}$.}
    \caption{Design Procedure for FSC-PA Decoder.}
\label{algo:sc-pa}
\end{algorithm}
% The schematic graph for a polar decoder with $N=8$ is depicted in Fig.~\ref{fig:decoderN8}. 
%  \begin{figure}[t]
%      \centering
%      \includegraphics[width=0.8\linewidth]{figures/decoderN8.eps}
%      \caption{SC Decoder for $N=8$}
%      \label{fig:decoderN8}
%  \end{figure}

\subsection{Proposed Design of FSC-PA Decoder}
The design of the \gls{FSC-PA} decoder depends on two parameters, the first is the maximum input size indicated by $n_0^{(0)}$, and the second is the inclusion threshold $\mathcal{P}_t$. The former can be found by estimating the decoding performance of the conventional \gls{EMS}-based decoder at different message sizes $n_m$. Once the good $n_m$ is found, it can be adopted in the design stage of the \gls{FSC-PA} decoder, i.e., $n_0^{(0)}=n_m$. 

\begin{figure}[t]
    \centering
    \includegraphics[width=\columnwidth]{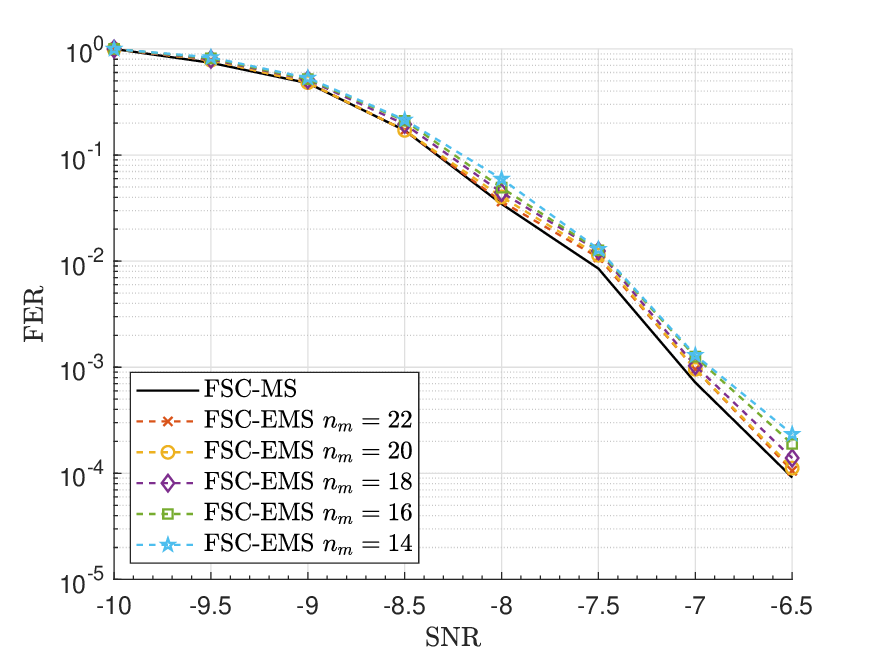}
    \caption{FER Performance of EMS-based SC Decoder for $K=42$ and $N=64$ over Different $n_m$.}
    \label{fig:compare_ems_nm}
\end{figure}

As an example, let the desired code be $N=64$ with $K=42$ on GF$(64)$ using \gls{CCSK} modulation. The first requirement is to find the decoding performance for different $n_m$ using Monte Carlo simulation (or a mathematical approximation as in \cite{nbpolar_ccsk}). The decoding performance of the \gls{EMS}-based decoder at different sizes $n_m$ is simulated as shown in Fig.~\ref{fig:compare_ems_nm} using \gls{CCSK} modulation and over an \gls{AWGN} channel. As shown, the parameter $n_m=18$ is the minimum size that maintains good performance at \gls{FER} $\approx 10^{-4}$. Hence, the parameter $n_0^{(0)}$ is set to 18 for the statistical analysis carried out in a \gls{FER} region of $10^{-2}$, that is, at an \gls{SNR} of $-7.5$ dB. 

Similarly, the parameter $n_0^{(0)}$ is deduced for other codes using the same approach. The value $n_0^{(0)}=18$ is found to be the minimum size that maintains good performance at \gls{FER} $\approx 10^{-4}$ for all code lengths $N \leq 512$.

Once the contribution rate matrices $C^{(l)}_s$ are obtained, the pruning process is performed. The threshold value $\mathcal{P}_t$ impacts both the complexity and the performance such that choosing a high threshold reduces the complexity and the decoding performance, and vice versa.

\begin{figure}[t]
    \centering
    \includegraphics[width=\columnwidth]{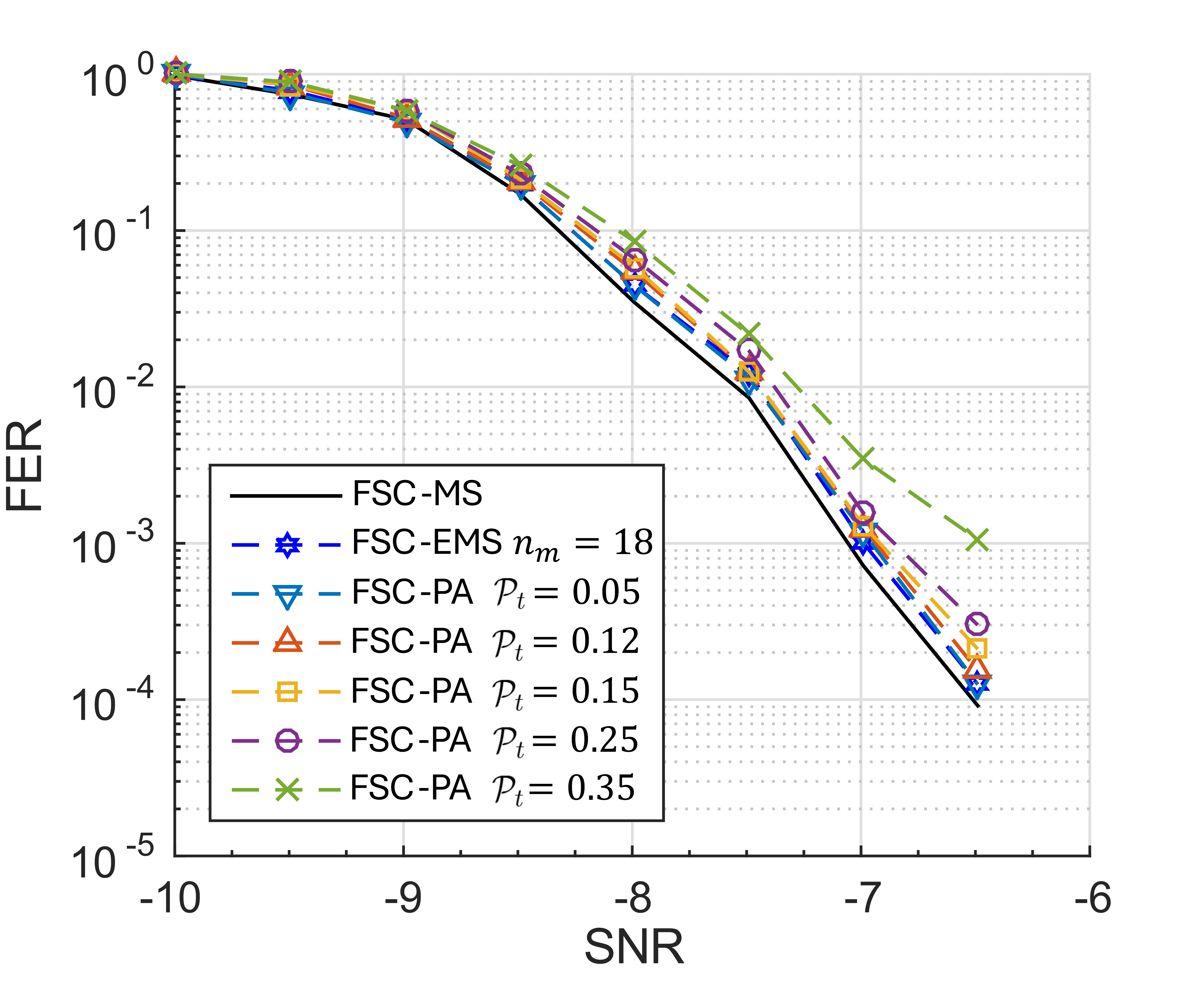}
    \caption{Performance of FSC-PA Decoder over Different Threshold Values $\mathcal{P}_t$ for $N=64$ and $K=42$.}
    \label{fig:compare_pa_pt}
\end{figure}

In Fig.~\ref{fig:compare_pa_pt}, the \gls{FER} performance of the proposed \gls{FSC-PA} decoder over different threshold values is depicted for a code of size $N=64$ and $K=42$ on GF(64) with \gls{CCSK} and over an \gls{AWGN} channel. The decoding performance is negligibly degraded compared to the FSC-MS for a threshold between $\mathcal{P}_t=0.05$ and $\mathcal{P}_t=0.12$. At $\mathcal{P}_t\geq0.15$, the degradation increases to $0.25$ dB. Hence, the threshold $\mathcal{P}_t=0.12$ is considered.

%The threshold value $\mathcal{P}_t$ is found using \gls{FER} comparison with the FSC-EMS decoder at the given $n_m$. The threshold value for $N=64$ is found as $\mathcal{P}_t=0.12$, and $\mathcal{P}_t=0.08$ for $N=256$. 
The design parameters $n_0^{(0)}$ and $\mathcal{P}_t$ have been found for different codes, as shown in table \ref{tab:design_parameter}. The parameter $r$ corresponds to the coding rate and is deduced as $r=K/N$. However, the association of \gls{CCSK} modulation spreads the sequence by a factor $S_f=p/q$, with $p = \log_2(q)$ denoting the number of bits associated with a \gls{CCSK} sequence of length $q$ ($S_f = 6/64$ for $q=64$). Therefore, the effective code rate is expressed as $r_e=rS_f$. The adopted threshold for complexity analysis and performance simulation is $\mathcal{P}_t=0.12$ for $N=64$ and $\mathcal{P}_t=0.08$ for $N=256$, and a message size $n_0^{(0)}=18$. 

\section{Complexity Analysis of FSC-PA Decoders}\label{sec:complexity}

 The complexity of the proposed \gls{FSC-PA} decoder is compared with the FSC-EMS decoder using a message size of $n_m$, the FSC-AEMS decoder with message sizes $n_L = 8$ and $n_H = 20$ \cite{sc_aems}, and the binary \gls{CA-FSCL32} decoder. A conclusive complexity analysis can be drawn when comparing the number of arithmetic and field operations across the \gls{NB} \gls{SC} decoders. However, this comparison becomes less straightforward when including the binary \gls{CA-FSCL32} decoder due to intricate sorting procedures, which are highly dependent on the decoder's structure and implementation, and cannot be accurately estimated. To address this, the number of arithmetic operations is considered to provide a reasonable insight into the computational complexity. Thus, shows whether real implementation is worth or not.
 
 Three types of constituent codes have been considered for \gls{NB}-\gls{SC} decoders \cite{fast_sc_nb}: Rate-0 codes, Rate-1 codes and REP codes of size $N_v=4$. In the binary \gls{CA-FSCL32} decoder, a fourth constituent code type, single parity check (SPC), is incorporated to improve the decoding efficiency. However, this constituent code is not used in the \gls{NB} decoders due to its high computational complexity in the \gls{NB} domain \cite{fast_sc_nb}. Furthermore, in the proposed \gls{FSC-PA} decoder, the final layers involve a limited processing, resulting in a minimal computational load. As a result, incorporating the SPC constituent code in this context would offer negligible additional complexity reduction. For further details on fast decoding, the reader is referred to \cite{fast_decoder,gen_fast_decoder,fast_sc_nb}. 
 
 However, it is worth noting that, in any \gls{NB}-\gls{SC} decoder, for Rate-1 constituent codes, the $N_v$ propagated messages consist of a single element, the hard decision. In the case of Rate-0 codes, recursive processing is completely bypassed, as all associated nodes correspond to frozen bits and are assigned a predetermined symbol, typically zero. For repetition (REP) codes, all $N_v$ nodes represent the same transmitted symbol; therefore, the $N_v$ \gls{LLR} input vectors each of size $q$ elements are aggregated through summation, and the symbol with the smallest \gls{LLR} value is decided.

 The complexity of a binary punctured \cite{punctured_pc} \gls{CA-FSCL32} decoder is evaluated for $K_b=Kp$ and $N_b=Np$. The complexity of the \gls{CA-FSCL32} decoder is assumed to be 32 more complex than that of the fast binary \gls{SC} decoder. It is clearly an underestimate of the complexity, since \gls{CA-FSCL32} requires many additional sorting operations between competing solutions. Without fast decoding, a binary \gls{SC} decoder performs a total of $\frac{N^{'}}{2} \log_2(N^{'})$ real additions with $N' = 2^{\lceil \log_2 N_b \rceil}$. With fast decoding, the $N_v/2$ additions and comparisons performed by the conventional code are omitted when rate-0 and rate-1 codes are substituted, replaced by $N_v$ additions for REP code, and replaced by $N_v$ comparators when SPC is substituted. The total number of additions performed by a \gls{CA-FSCL32} decoder is provided in Table \ref{tab:design_parameter}.

Alternatively, in \gls{NB} FSC-EMS decoders using the bubble check approach \cite{bubble_check}, a \gls{CN} that generates an output of size $n_m$ requires $O^\oplus_{\text{EMS}}$ \gls{GF} additions and $O^+_{\text{EMS}}$ \gls{LLR} additions, which are given by:
\begin{equation}
    \begin{gathered}
        O^\oplus_{\text{EMS}}=n_m\sqrt{n_m}; \
        O^+_{\text{EMS}}=n_m\sqrt{n_m}-2n_m+1.
    \end{gathered}
    \label{eq:nb_op_ems_node}
\end{equation}

Furthermore, an \gls{AEMS}-based \gls{CN} requires candidates $O^\oplus_{\text{AEMS}}$ \gls{GF} and candidates $O^+_{\text{AEMS}}$ \gls{LLR} estimated as follows.
\begin{equation}
    \begin{gathered}
        O^\oplus_{\text{AEMS}}=2(n_L+n_H)-4;\ 
        O^+_{\text{AEMS}}=n_L+n_H-3.
    \end{gathered}
    \label{eq:nb_op_aems_node}
\end{equation}

On the other hand, an \gls{FSC-PA} \gls{CN} has a unique intermediate matrix $T'^{(l)}_{\Sigma_s}$ depending on its corresponding layer $l$ and cluster $s$. Therefore, the numbers of operations are: $O^{l\oplus}_{s_{\text{PA}}}$ \gls{GF} additions and $O^{l+}_{s_{\text{PA}}}$ \gls{LLR} candidates that can be expressed as: 

\begin{equation}
    \begin{gathered}
    O^{l\oplus}_{s_{\text{PA}}}=\sum_{i=0}^{n^{(l)}_{s}-1}\sum_{j=0}^{n^{(l)}_{s}-1}{\mathcal{R}^{(l)}_{s}(i,j)};\
    O^{l+}_{s_{\text{PA}}}=\sum_{i=1}^{n^{(l)}_{s}-1}\sum_{j=1}^{n^{(l)}_{s}-1}{\mathcal{R}^{(l)}_{s}(i,j)}.
    \end{gathered}
    \label{eq:pa_node}
\end{equation}

It is noticeable that the number of real additions is less than the field additions in all estimations. This is because the processing of the first row and column of any intermediate matrix ($T'_{\Sigma}$\text{ or }$T'^{(l)}_{\Sigma_s}$) requires only \gls{GF} additions (the \gls{LLR} value of $M_H(0)$ and $M_L(0)$ is always zero due to normalisation).

 The total number of \gls{GF} (denoted as $T^\oplus$) and \gls{LLR} (denoted as $T^+$) additions performed by the FSC decoder (after applying fast decoding) is shown in table \ref{tab:design_parameter} for each code length $N$ and rate $r$.

  For the \glspl{VN} processing, no optimisation or simplification is assumed for all \gls{NB} decoders for ease of evaluation. A \gls{VN} performs the processing in \eqref{eq:vn_update_ms}, thus requiring $q$ real additions. Thus, without fast decoding, the total real additions required by the \glspl{VN} in an \gls{NB}-\gls{SC} decoder are $\frac{N}{2} \log_2(N)q$ additions. However, each rate-0 and rate-1 code saves $N_vq/2$ additions, while REP constituent codes replace all the \glspl{CN} and \glspl{VN} arithmetic and logical operations by $N_vq$ additions only. Table \ref{tab:design_parameter} provides the total number of arithmetic operations (GF and real) for different coding rates and lengths.

\begin{table*}[t]
    \centering
    \caption{Design Parameters for FSC-PA Decoder and Required Addition Operations over Different Code lengths.}
    \begin{tabular}{|c|c|c|c|c|c|c|c|c|c|c|c|c|c|}
        \hline
        \multirow{3}{*}{$K$}&\multirow{3}{*}{$N$}&\multirow{3}{*}{$r$}&\multirow{3}{*}{$r_e$}& \multirow{2}{*}{\gls{CA-FSCL32}} & \multicolumn{2}{|c|}{FSC-EMS} & \multicolumn{2}{|c|}{FSC-AEMS} & \multicolumn{3}{|c|}{FSC-PA}\\
        \cline{6-12}
        &&&&&\multicolumn{2}{|c|}{$n_m=18$}&\multicolumn{2}{|c|}{$n_L=20, n_H=8$}&\multicolumn{3}{|c|}{$\mathcal{P}_t=0.12,\ n_0^{(0)}=18$}\\
        \cline{5-12}
        &&&&$T^+$&$T^\oplus$&$T^+$&$T^\oplus$&$T^+$&$T^\oplus$&$T^+$&$\mu$\\
        \hline
        11&\multirow{3}{*}{64}&0.17&0.016&35456&5036&6603&3432&5874&1962&4720&$-$13.5 dB\\
        21&&0.33&0.031&43392&7934&11313&5408&9256&2956&7236&$-$10.5 dB\\
        42&&0.66&0.062&46464&8543&12697&5824&9968&3224&7520&$-$7.5 dB\\
        \hline
        \multicolumn{5}{|c|}{}&\multicolumn{2}{|c|}{$n_m=18$}&\multicolumn{2}{|c|}{$n_L=20, n_H=8$}&\multicolumn{3}{|c|}{$\mathcal{P}_t=0.08,\ n_0^{(0)}=18$}\\
        \hline
        43&\multirow{3}{*}{256}&0.16&0.016&203840&24729&33801&16848&28836&10094&23442&$-$14 dB\\ 
        85&&0.33&0.031&189824&29003&41341&19760&33820&11966&26990&$-$11.5 dB\\
        171&&0.67&0.063&220032&40757&60560&27768&47526&15758&36110&$-$8 dB\\
        \hline
    \end{tabular}
    \label{tab:design_parameter}
\end{table*}

 Based on the aforementioned complexity analysis, the \gls{FSC-PA} decoder saves at least $60\%$ of the computation resources for field additions and around $30\%$ of the real additions compared to the FSC-EMS decoder. Compared to the FSC-AEMS decoder, the \gls{FSC-PA} decoder saves at least $40\%$ of the computation resources for field additions and around $20\%$ of the real additions. This assessment disregards any complexity savings of \glspl{VN} and the sorting process performed at \glspl{CN}. However, the \gls{FSC-PA} decoder requires additional comparators for input reallocation as in \eqref{eq:relocate_inputs}.
 
 Compared to the binary \gls{CA-FSCL32} decoder, the proposed \gls{FSC-PA} decoder requires only 10\% of the total real additions. Moreover, the complexity of the \gls{FSC-PA} decoder is believed to be further reduced when observing the \glspl{VN} statistics independently. Consequently, further complexity investigation based on implementation results is worth.

 % Furthermore, an analytical evaluation has been performed for $K=21$ and $N=64$ symbols. The assessment determines the number of arithmetic (\gls{GF} and \gls{LLR} additions) and logical (comparison) operations performed by the proposed \gls{FSC-PA} decoder with $n_m=25$ and $\mathcal{P}_t=0.07$, and the binary \gls{CA-FSCL32} decoder with a list size of $L=32$. The results obtained show that the \gls{CA-FSCL32} decoder requires 38464 comparisons and 42560 additions, while the \gls{FSC-PA} decoder requires 3436 \gls{GF} additions, 7616 \gls{LLR} additions, and 14048 comparisons. Considering that a 6-bit $(\log_2(q))$ adder is twice as complex as a 6-bit comparator, and that the complexity of a 6-bit comparator is equivalent to 20 XOR gates, the proposed \gls{FSC-PA} decoder with $n_m=25$ and $\mathcal{P}_t=0.07$ is four times less complex than the \gls{CA-FSCL32} decoder.

\section{Simulation Results and Performance Analysis}
The decoding performance is evaluated using Monte Carlo simulations over an \gls{AWGN} channel. The \gls{FER} of the FSC-MS, FSC-EMS, and \gls{FSC-PA} decoders is simulated and compared under \gls{CCSK} \cite{ccsk} and \gls{BPSK} modulation. Furthermore, the performance of the binary punctured \cite{punctured_pc} \gls{CA-FSCL32} decoder is assessed and a fixed CRC polynomial defined in hexadecimal form as 0x1864CFB is used consistently in all simulations.

The coefficients $\gamma$ of all \gls{NB} decoders over \gls{CCSK} are set to 1 across all layers, following the findings of \cite{nbpolar_ccsk_coeff}, and are determined as in \cite{nbpolar_ccsk} over \gls{BPSK} modulation.

\begin{figure}[t]
    \centering
    \includegraphics[width=\columnwidth, height = 6.8cm]{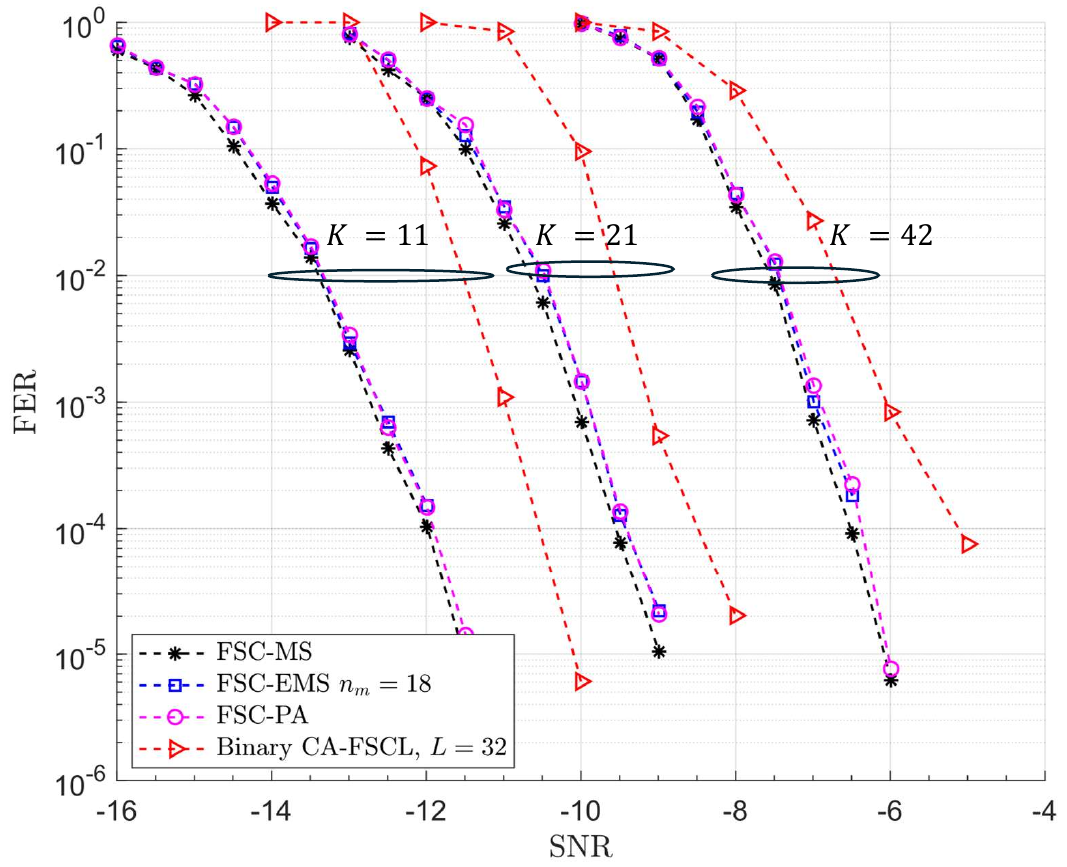}
    \caption{Simulation Results for $N=64$ over GF(64).}
    \label{fig:sim64}
\end{figure}

Figure~\ref{fig:sim64} presents the decoding performance of FSC-MS, FSC-EMS, and \gls{FSC-PA} decoders for a code length of $N=64$ symbols over GF$(64)$ and information lengths of $K=11$, $21$, and $42$ symbols. The decoders are configured according to the parameters listed in Table~\ref{tab:design_parameter}. The performance of the binary \gls{CA-FSCL32} decoder is also included for $K_b=Kp$ and $N_b=Np$ bits, where $p=\log_2(q)$. As observed, the binary \gls{CA-FSCL32} decoder experiences performance degradation compared to its \gls{NB} counterparts, by approximately 1.7~dB at low rates and 1~dB at high rates, due to binary marginalisation, which is inherently avoided when using \gls{NB} codes with the same modulation order. The \gls{FSC-PA} decoder exhibits nearly identical performance to the FSC-EMS decoder and diverges slightly, by about 0.15~dB, from the FSC-MS decoder.

\begin{figure}[t]
    \centering
    \includegraphics[width=\columnwidth]{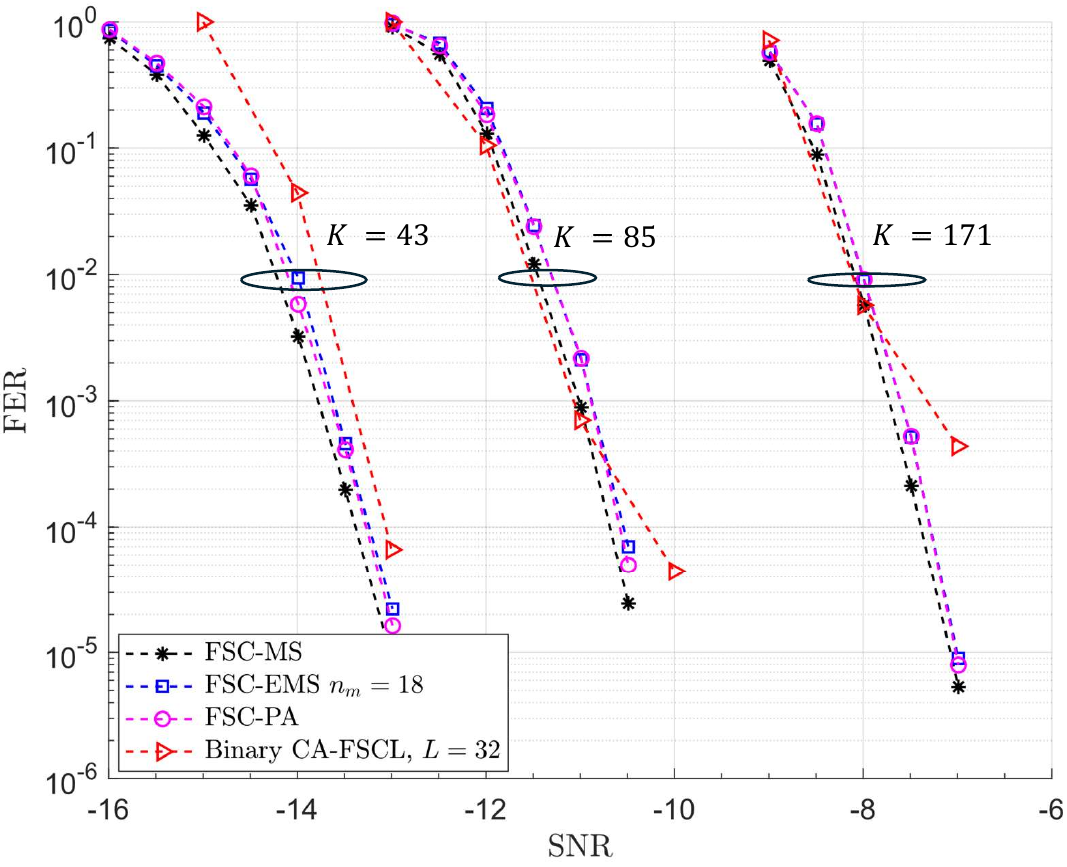}
    \caption{Simulation Results for $N=256$ over GF(64).}
    \label{fig:sim256}
\end{figure}

Figure~\ref{fig:sim256} illustrates the decoding performance for $N=256$ and $K=43$, $85$, and $171$ symbols on GF$(64)$ with \gls{CCSK} modulation. The performance of binary \gls{CA-FSCL32} is also shown in red, using $K_b=Kp$ and $N_b=Np$ bits. Compared to the \gls{NB} polar decoders, the binary \gls{CA-FSCL32} exhibits a performance degradation of approximately 0.4~dB at low rates, while achieving comparable performance at higher rates down to a \gls{FER} of $10^{-3}$. Nonetheless, the binary decoder suffers from an earlier error floor at lower \glspl{FER}.

\begin{figure}[t]
    \centering
    \includegraphics[width=\columnwidth]{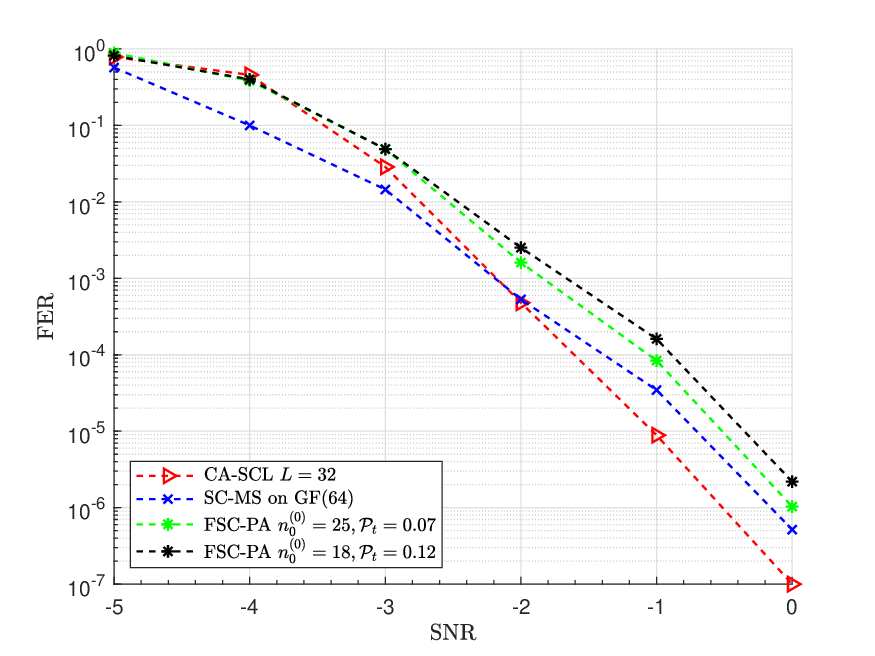}
    \caption{BPSK Simulation Results for $N=64$ and $K=21$.}
    \label{fig:bpsk_sim}
\end{figure}

Fig.~\ref{fig:bpsk_sim} compares \gls{NB} \glspl{PC} and binary \glspl{PC} under \gls{BPSK} modulation. The binary code is a punctured \gls{PC} of length 384 bits, with 126 information bits, while the \gls{NB}-\gls{PC} is defined over GF(64), with $N=64$ and $K=21$. Two sets of parameters are used for the \gls{FSC-PA} decoder: the first, given in Table~\ref{tab:design_parameter} uses $n^{(0)}_0=18$ and $\mathcal{P}_t = 0.12$); the second is selected to improve performance, with ($n^{(0)}_0=25$, $\mathcal{P}_t = 0.07$). Even with these improved parameters, the \gls{CA-FSCL32} decoder consistently outperforms the \gls{FSC-PA}, exhibiting a steeper error-rate slope. However, this gain in performance comes at the expense of higher computational complexity: $T^+ = 43392$ real additions for \gls{CA-FSCL32}) against $T^+=8554$ for the \gls{FSC-PA} decoder with parameters $n_m=25$ and $\mathcal{P}_t = 0.07$). 
These results call for a last comments: for \glspl{FER} above $10^{-3}$, the FSC-MS decoder outperforms the \gls{CA-FSCL32}. This means that if cost is not an issue, the \gls{NB}-\gls{PC} may be of interest due to its performance at \glspl{FER} above $10^{-3}$. This point is further reinforced by the fact that a list decoder can also be applied to the \gls{NB}-\gls{PC} (this point lies beyond the scope of this paper).

\section{Conclusion}\label{sec:conc}
This paper introduces an \gls{FSC-PA} decoder for \gls{NB}-\glspl{PC}, offering a simplified yet effective approach to reducing decoding complexity through per node clustering dedicated check node optimisation. The proposed \gls{FSC-PA} decoder achieves a 60 \% reduction in field additions and a 30 \% reduction in real additions compared to the FSC-EMS decoder for codes over GF(64). Under \gls{CCSK} modulation (a quasi-orthogonal modulation scheme) the proposed decoder outperforms the state-of-the-art binary \gls{CA-FSCL32} decoder by up to 1.7 dB, while maintaining significantly lower complexity. In contrast, under \gls{BPSK} modulation, the \gls{CA-FSCL32} retains a performance advantage, albeit at the cost of higher computational complexity. The ability of the \gls{FSC-PA} decoder to surpass \gls{CA-FSCL32} under CCSK modulation (both in decoding performance and complexity) highlights its potential as a promising solution for low power wide area network. %The combination of \gls{FSC-PA} with orthogonal modulation also enables robust frame detection and synchronisation at very low SNRs, further strengthening its suitability for short-packet transmission in challenging environments

\bibliography{IEEEabrv,main}

\begin{thebibliography}{10}
\providecommand{\url}[1]{#1}
\csname url@samestyle\endcsname
\providecommand{\newblock}{\relax}
\providecommand{\bibinfo}[2]{#2}
\providecommand{\BIBentrySTDinterwordspacing}{\spaceskip=0pt\relax}
\providecommand{\BIBentryALTinterwordstretchfactor}{4}
\providecommand{\BIBentryALTinterwordspacing}{\spaceskip=\fontdimen2\font plus
\BIBentryALTinterwordstretchfactor\fontdimen3\font minus
  \fontdimen4\font\relax}
\providecommand{\BIBforeignlanguage}[2]{{%
\expandafter\ifx\csname l@#1\endcsname\relax
\typeout{** WARNING: IEEEtran.bst: No hyphenation pattern has been}%
\typeout{** loaded for the language `#1'. Using the pattern for}%
\typeout{** the default language instead.}%
\else
\language=\csname l@#1\endcsname
\fi
#2}}
\providecommand{\BIBdecl}{\relax}
\BIBdecl

\bibitem{Liva17}
G.~Liva, B.~Matuz, E.~Paolini, and M.~F. Flanagan, ``{Non-binary LDPC codes for
  orthogonal modulations: Analysis and code design},'' in \emph{2017 IEEE Int.
  Conf. Commun (ICC)}, 2017, pp. 1--6.

\bibitem{Bourduge23}
J.~Bourduge, C.~Poulliat, and B.~Gadat, ``{Non Binary LDPC Coded Orthogonal
  Modulation Schemes based on Non Binary Multilevel Coding},'' in \emph{2023
  IEEE Int. Symp. Inf. Theory (ISIT)}, 2023, pp. 2517--2522.

\bibitem{arikanpc}
E.~Arikan, ``{Channel Polarization: A Method for Constructing
  Capacity-Achieving Codes for Symmetric Binary-Input Memoryless Channels},''
  \emph{IEEE Trans. Inf. Theory}, 2009.

\bibitem{ccsk_nbcodes}
O.~Abassi, L.~Conde-Canencia, M.~Mansour, and E.~Boutillon, ``{Non-Binary Coded
  CCSK and Frequency-Domain Equalization with Simplified LLR Generation},'' in
  \emph{2013 IEEE 24th Int. Symp. Pers. Indoor Mob. Radio Commun (PIMRC)},
  2013, pp. 1478--1483.

\bibitem{ks_ccsk}
K.~Saied, A.~C.~A. Ghouwayel, and E.~Boutillon, ``{Short Frame Transmission at
  Very Low SNR by Associating CCSK Modulation With NB-Code},'' \emph{IEEE
  Trans. Wireless Commun.}, vol.~21, no.~9, pp. 7194--7206, 2022.

\bibitem{LoRa}
F.~Sforza, ``Communications system,'' U.S. Patent US8\,406\,275B2, 2013.

\bibitem{fast_decoder}
G.~Sarkis, P.~Giard, A.~Vardy, C.~Thibeault, and W.~J. Gross, ``Fast polar
  decoders: Algorithm and implementation,'' \emph{IEEE J. Sel. Areas Commun.},
  vol.~32, no.~5, pp. 946--957, 2014.

\bibitem{gen_fast_decoder}
C.~Condo, V.~Bioglio, and I.~Land, ``Generalized fast decoding of polar
  codes,'' in \emph{2018 IEEE Global Commun Conf. (GLOBECOM)}, 2018, pp. 1--6.

\bibitem{nbpolar_ccsk}
V.~Savin, ``{Non-Binary Polar Codes for Spread-Spectrum Modulations},'' in
  \emph{2021 11th Int. Symp. on Topics in Coding (ISTC)}, 2021, pp. 1--5.

\bibitem{sc_ms}
F.~Cochachin, L.~Luzzi, and F.~Ghaffari, ``{Reduced Complexity of a Successive
  Cancellation Based Decoder for NB-Polar Codes},'' in \emph{2021 11th Int.
  Symp. on Topics in Coding (ISTC)}, 2021, pp. 1--5.

\bibitem{nbpolar_ccsk_coeff}
F.~Cochachin and F.~Ghaffari, ``{A Lightweight Encoder and Decoder for
  Non-Binary Polar Codes},'' in \emph{2023 Congr. Comput. Sci. Comput. Eng.
  Appl. Comput. (CSCE)}, 2023, pp. 979--984.

\bibitem{sc_ems}
\BIBentryALTinterwordspacing
P.~Chen, B.~Bai, and X.~Ma, ``{Non-Binary Polar Coding with Low Decoding
  Latency and Complexity},'' \emph{Journal of Information and Intelligence},
  2022. [Online]. Available:
  \url{https://www.sciencedirect.com/science/article/pii/S294971592200004X}
\BIBentrySTDinterwordspacing

\bibitem{ems_ldpc}
D.~Declercq and M.~Fossorier, ``{Extended Min-Sum Algorithm for Decoding LDPC
  Codes Over GF(q)},'' in \emph{Proc.. Int. Symp. Inf. Theory, 2005. ISIT
  2005.}, 2005, pp. 464--468.

\bibitem{sc_aems}
J.~Jabour, A.~C. Al~Ghouwayel, and E.~Boutillon, ``{Asymmetrical Extended
  Min-Sum for Successive Cancellation Decoding of Non-Binary Polar Codes},'' in
  \emph{2023 13th Int. Symp. on Topics in Coding (ISTC)}, 2023.

\bibitem{l_bubble}
E.~Boutillon and L.~Conde-Canencia, ``{Simplified Check Node Processing in
  Non-Binary LDPC Decoders},'' in \emph{2010 6th Int. Symp. on Turbo Codes
  Iterative Inf. Processing}, 2010, pp. 201--205.

\bibitem{ccsk}
G.~Dillard, M.~Reuter, J.~Zeiddler, and B.~Zeidler, ``{Cyclic Code-Shift
  Keying: A Low Probability of Intercept Communication Technique},'' \emph{IEEE
  Trans. Aerosp. Electron. Syst.}, vol.~39, no.~3, pp. 786--798, 2003.

\bibitem{rathi2005}
V.~Rathi and R.~Urbanke, ``Density evolution, thresholds and the stability
  condition for non-binary ldpc codes,'' \emph{IEE Proceedings-Communications},
  vol. 152, no.~6, pp. 1069--1074, 2005.

\bibitem{fast_sc_nb}
A.~Farsiabi, H.~Ebrahimzad, M.~Ardakani, and C.~Li, ``{Fast
  Successive-Cancellation Decoding of 2 × 2 Kernel Non-Binary Polar Codes},''
  \emph{IEEE Trans. Commun.}, vol.~73, no.~2, pp. 1009--1024, 2025.

\bibitem{punctured_pc}
V.~Bioglio, F.~Gabry, and I.~Land, ``Low-complexity puncturing and shortening
  of polar codes,'' in \emph{2017 IEEE Wirel. Comm. Netw. Conf. Workshops
  (WCNCW)}, 2017, pp. 1--6.

\bibitem{bubble_check}
E.~Boutillon and L.~Conde-Canencia, ``{Bubble check: a simplified algorithm for
  elementary check node processing in Extended Min-Sum non-binary LDPC
  decoders},'' \emph{IEEE Electron. Lett.}, vol.~46, no.~9, pp. 633--631, 2010.

\end{thebibliography}
\end{document}